\documentclass[twoside,letterpaper]{soups} 
\usepackage{graphicx}

\usepackage{color}
\usepackage[hyphens]{url}

\usepackage{listings}
\usepackage{booktabs}

\newcommand{\eps}{$\varepsilon$}

\usepackage[table]{xcolor}

\conferenceinfo{}{}
\CopyrightYear{2014}

\global\boilerplate={This work is licensed under the Creative Commons Attribution 4.0 International License. To view a copy of this license, visit http://creativecommons.org/licenses/by/4.0/ or send a letter to Creative Commons, 444 Castro Street, Suite 900, Mountain View, California, 94041, USA.}

\begin{document}
\title{Through the Frosted Glass:\\Security Problems in a Translucent UI}
\subtitle{v1}
\numberofauthors{4}
\author{
\alignauthor Arne Renkema-Padmos\\
  \affaddr{CASED / TU Darmstadt}\\
  \affaddr{Hochschulstra\ss e 10, 64289, Darmstadt, Germany}\\
  \email{arne.renkema-padmos@cased.de}
\and
\alignauthor Jerome Baum\\
  \affaddr{Independent Researcher}\\
  \affaddr{Narzissenweg 26, 37081 G\"ottingen}\\
  \email{jerome@jeromebaum.com}
}


\maketitle

\begin{abstract}
Translucency is now a common design element in at least one popular mobile operating system. This raises security concerns as it can make it harder for users to correctly identify and interpret trusted interaction elements. In this paper, we demonstrate this security problem using the example of the Safari browser in the latest iOS version on Apple tablets and phones (iOS7), and discuss technical challenges of an attack as well as solutions to these challenges. We conclude with a survey-based user study, where we seek to quantify the security impact, and find that further investigation is warranted.
\end{abstract}



\section{Introduction}


Graphic design has gone through many phases throughout history. Over the last century we have seen (among others) expressionism, avant garde, modernism, and post-modernism, followed by what could be called ``contemporary" design. User interface (UI) design can also be seen as consisting of several distinct visual styles over the field's history. Initially there were physical buttons and printers for output. This moved to terminals, where things went from command line, to single windows, to overlapping windows, and back to non-overlapping interfaces with new mobile interfaces. A distinct trend in recent years that can be observed is that of ``flat" design, whereby the fake shadows and crystal buttons of the Web 2.0 graphical movement have been replaced by something slimmed down, and less ``metaphoric". One place where this can be seen is in ``the great flattening" of the iOS7 interface from Apple.

\begin{figure}
\centering
\frame{\includegraphics[width=\columnwidth]{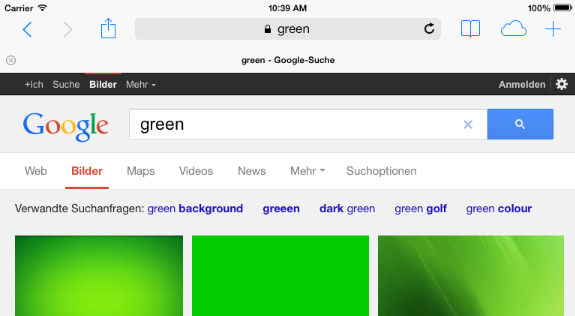}}
\mbox{}
\frame{\includegraphics[width=\columnwidth]{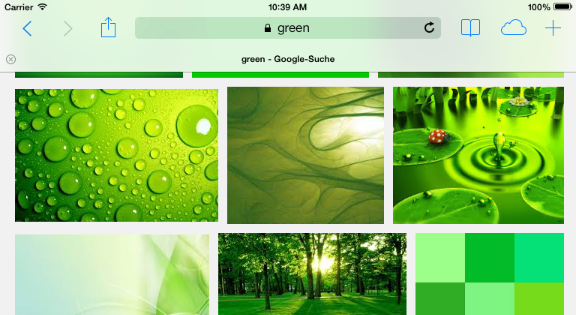}}
\caption{Demonstration of transparency in iOS7}
\label{fig:example-transparency}
\end{figure}

As graphical power has grown in recent years, Apple has added transparency effects to iOS7. This graphical power enabled, among others, the parallax effect and layered translucency. In the Safari iOS7 web browser this translucency is also implemented, as shown in Figure \ref{fig:example-transparency}.


In this paper we present an attack that may enhance phishing attacks (Section \ref{chap:attack}), which is based on this transparency effect. We tested our attack using a crowdsourced experiment, and find that a larger study or an improved attack with larger effect is needed to find conclusive evidence of a threat (Section \ref{chap:evaluation}). Based on some further exploratory work we provide proposals for the design of trusted user interfaces (Section \ref{chap:solution}). Our contributions are:

\begin{itemize}
\item identifying the potential problems that underlay attacks can cause,
\item evaluating the impact of such an attack in a major smartphone and tablet operating system,
\item running a usable security study on multiple crowdsourcing platforms.
\end{itemize}








\begin{figure}\label{fig:attack-screenshot}
\centering
\frame{\includegraphics[width=\columnwidth]{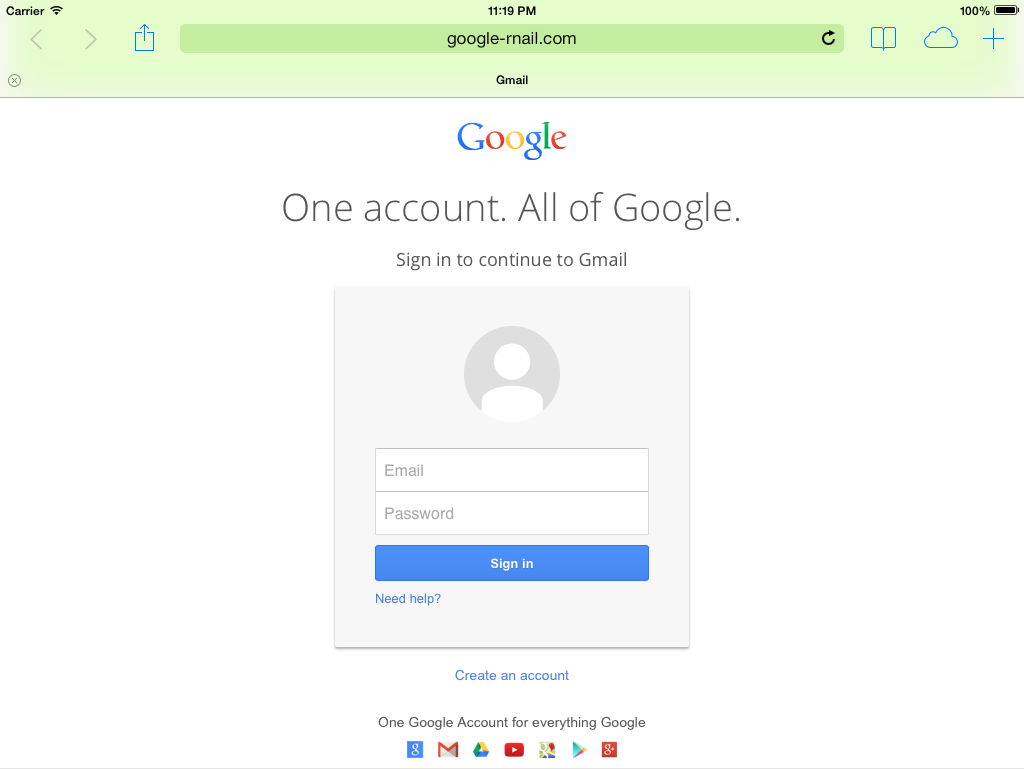}}
\caption{Example of an underlay attack}
\end{figure}

\begin{figure}\label{fig:attack-screenshot2}
\centering
\frame{\includegraphics[width=\columnwidth]{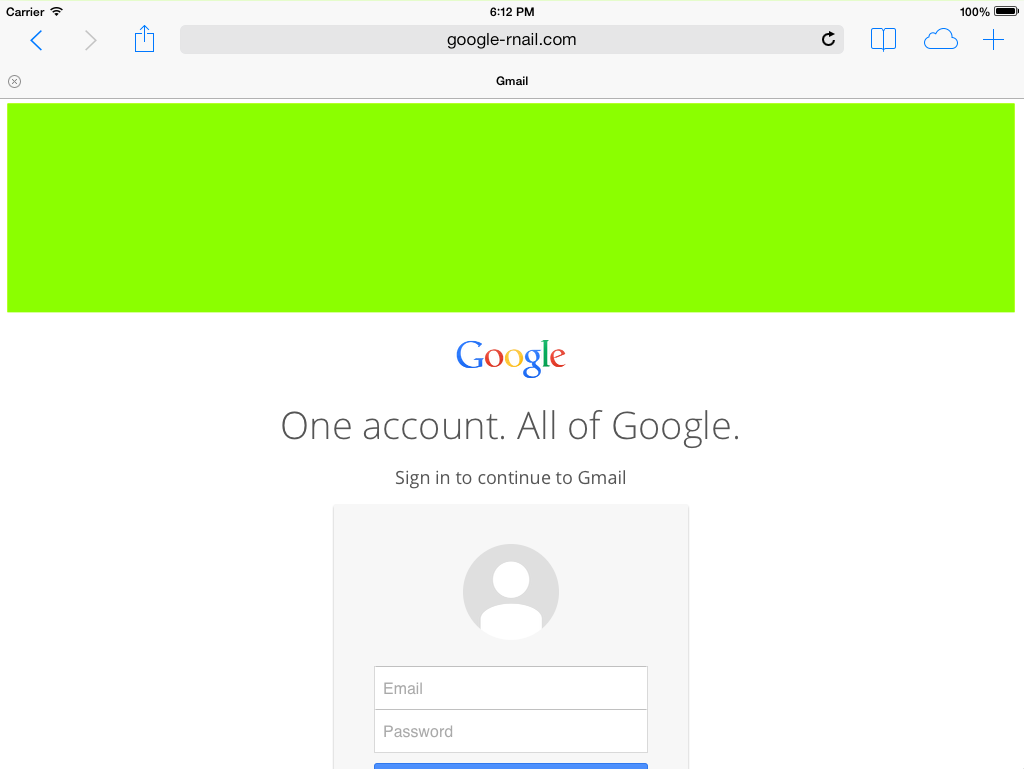}}
\caption{The underlay attack exposed}
\end{figure}

\begin{figure}\label{fig:attack-layers}
\centering
\frame{\includegraphics[width=\columnwidth]{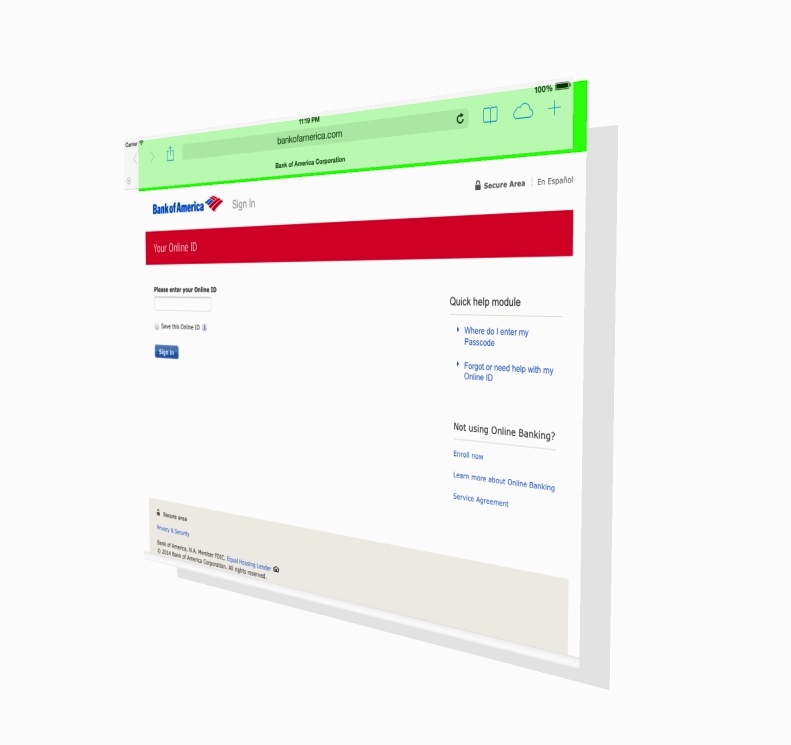}}
\caption{Illustration of the attack idea}
\end{figure}

\section{The Underlay Attack}\label{chap:attack}

\begin{figure}
\centering
\frame{\includegraphics[width=\columnwidth]{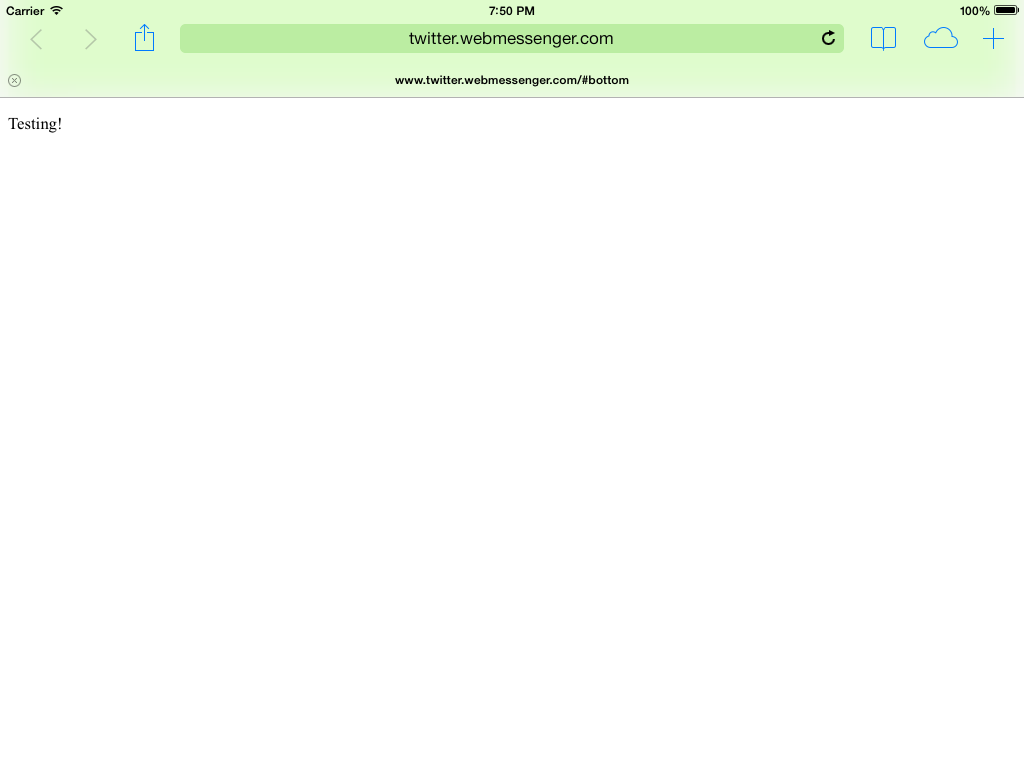}}
\caption{Our proof of concept for the attack}
\label{fig:proof-of-concept}
\end{figure}

Using the code in Figure \ref{fig:attack-code} we were able to create a proof of concept for the attack. The result is shown in Figure \ref{fig:proof-of-concept}. The code is specifically aimed at the iPad's landscape mode, and doesn't include a copy of a target website yet. For an actual attack, it would be trivial for the attacker to include the authentic website in an IFRAME, detect the orientation and respond dynamically, and integrate password capture mechanisms. Figure \ref{fig:attack-screenshot} shows an example of what a real attack could look like.


\begin{figure}\label{fig:attack-code}
\begin{lstlisting}
<style>
.green {
    width: 100%;
    min-height: 200px;
    background-color: #88ff00;
}
</style>
...
<div class="green">&nbsp;</div>
<div id="high">
    ...
</div>
<a name="login">&nbsp;</a>
<script>
document.getElementById('high').style.height = window.innerHeight - 45;
var anchor = "#login";
if (location.hash != anchor) {
    location.href = location.pathname + anchor;
}
</script>
\end{lstlisting}
\caption[]{Attack code (CSS, HTML, JavaScript)}
\end{figure}



Given that the colour green is widely used in secure user interfaces we expected that this would cause more people to fall for phishing. If the user were to scroll up in the page then the green bar would become visible and the deceipt would be obvious. However there are technical ways to prevent this from happening. The details of such an implementation are a technicality and therefore we did not concern ourselves with this for the purpose of this study.


Translucency can be a visually appealing design element. However, this may come at the cost of decreased security. Yee \cite{yee2005guidelines} provides guidelines for secure user interaction design. The paper states, among other things that interfaces should represent objects and actions using unspoofable representations. When an attacker-controlled UI element is presented underneath a trustworthy but transparent element, this gives the attacker limited control over the trustworthy element. Yee also states that authority relationships that would affect security-relevant decisions should be easily reviewable in the interface. From controlling a trustworthy UI element it is not a long way to spoofing system security marks. Attacks always get better, they never get worse.

Transparency is used throughout the user interface design in Apple's current iOS operating system for iPad tablets. The included web browser in iOS is Safari. One noteworthy instance of transparency design is the browser's address bar.

Using colouring in the address bar is an emerging trend for indicating SSL certificate information and other security aspects of a connected website. As illustrated in Figure \ref{fig:attack-layers}, the colour of Safari's address bar depends in part on the colour of the web page loaded underneath. An attack is then straightforward: direct the browser to a page that automatically scrolls, such that a big green-coloured area is placed directly underneath the address bar.

Besides trying to increase trustworthiness of the website though a coloured underlay, other avenues for attack may be possible, such as trying to make the URL harder to read by decreasing the contrast of the URL bar.

\begin{figure}
\centering
{
\begin{tabular}{rcc}
  & \textbf{Bank of America} & \textbf{Google Mail} \\\\
  & \emph{Authentic with EV-SSL} & \emph{Authentic with SSL}\\
  & \emph{bankofamerica.com} & \emph{accounts.google.com}\\
  & \frame{\includegraphics[width=0.2\textwidth]{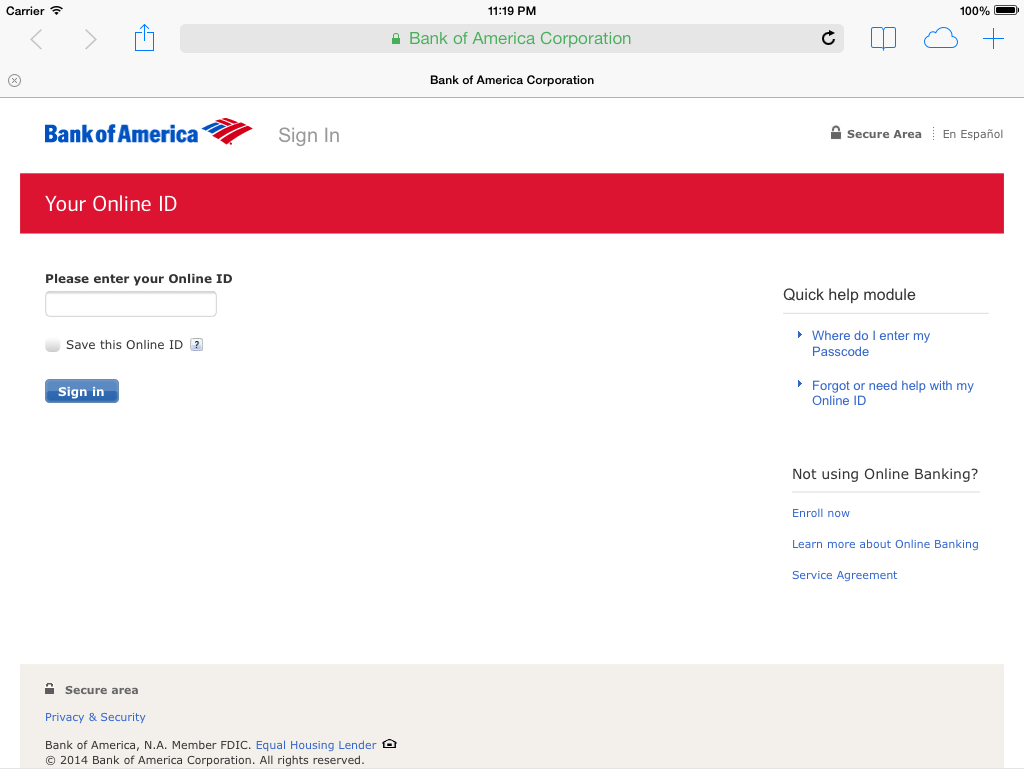}}
  & \frame{\includegraphics[width=0.2\textwidth]{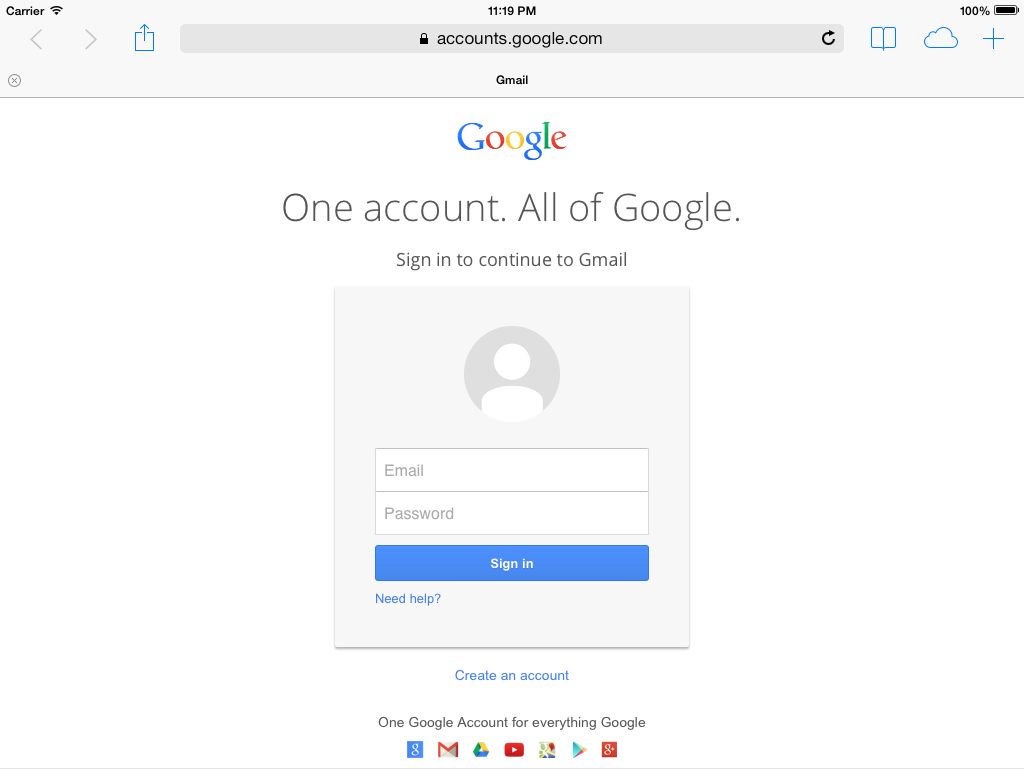}}\\\\
  & \emph{Fraud with green bar} & \emph{Fraud with green bar}\\
  & \emph{bankofarnerica.com} & \emph{google-rnail.com}\\
  & \frame{\includegraphics[width=0.2\textwidth]{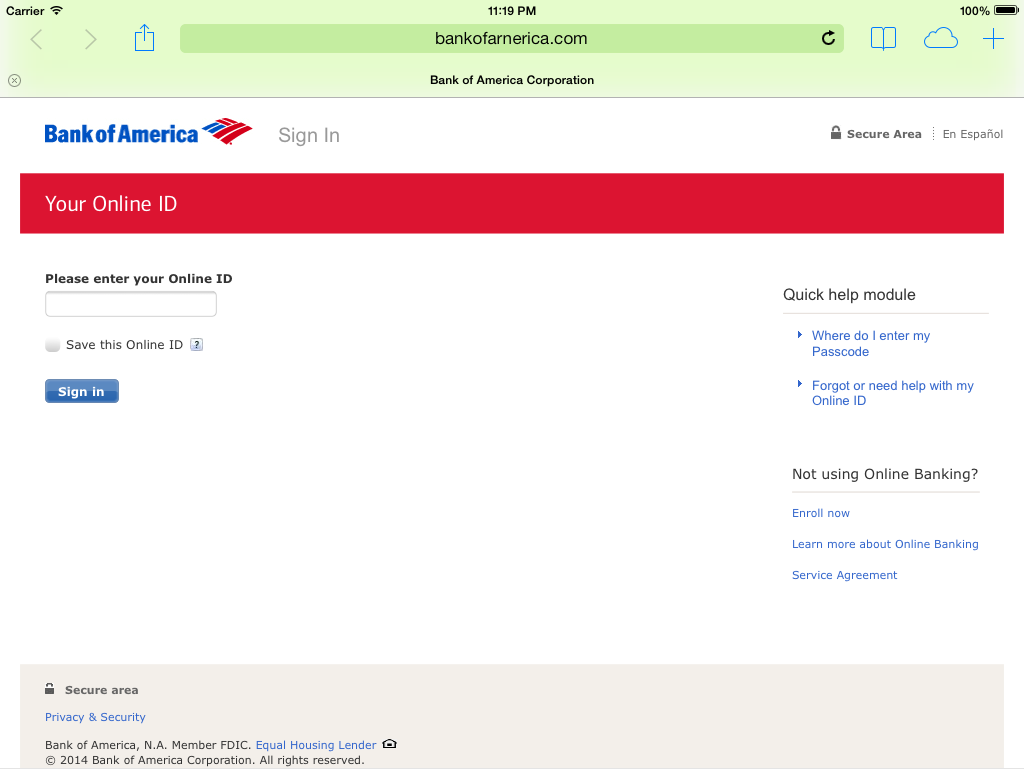}}
  & \frame{\includegraphics[width=0.2\textwidth]{survey-screenshots/gmail_montage_phishing_color}}\\\\
  & \emph{Fraud without green bar} & \emph{Fraud without green bar}\\
  & \emph{bankofarnerica.com} & \emph{google-rnail.com}\\
  & \frame{\includegraphics[width=0.2\textwidth]{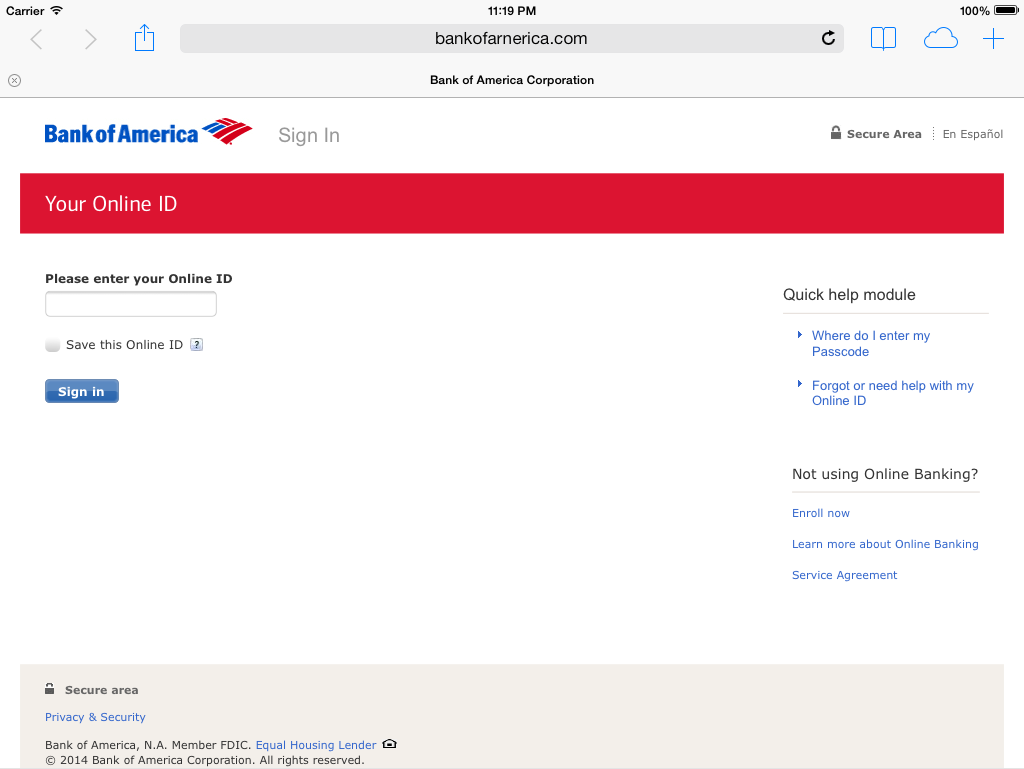}}
  & \frame{\includegraphics[width=0.2\textwidth]{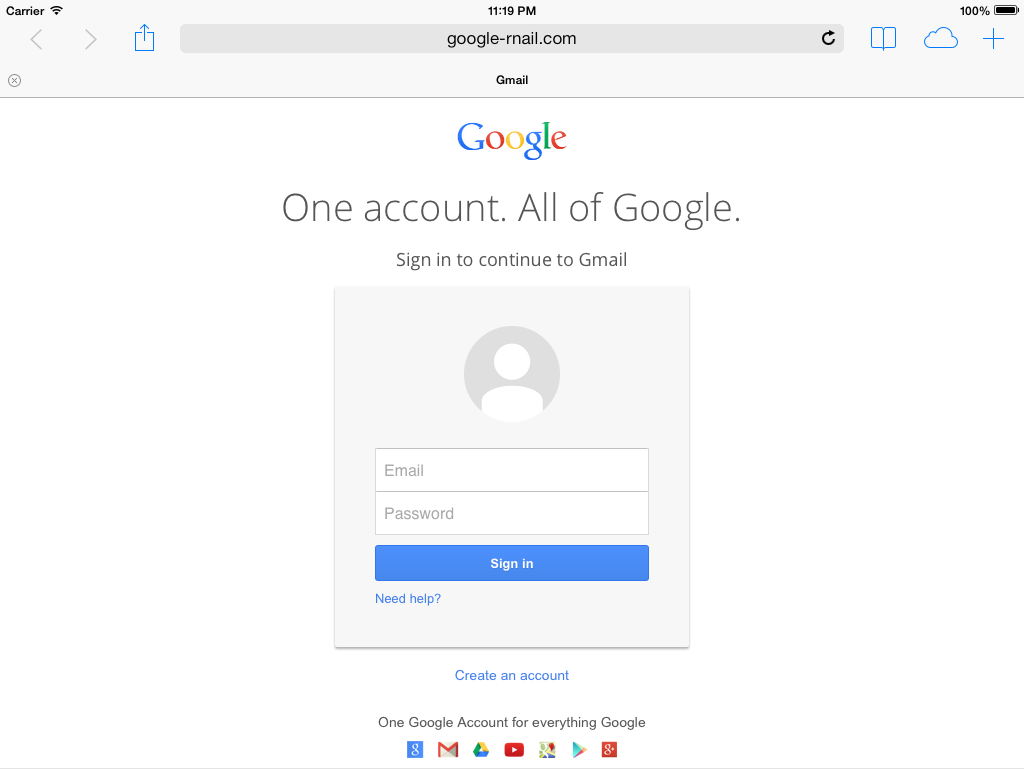}}
\end{tabular}
}
\caption{iPad screenshots used in the survey}
\label{fig:survey-screenshots}
\end{figure}

\section{Evaluation of Threat Level}\label{chap:evaluation}


Our research question was \emph{can subversion of the interface via translucent elements lead to negative security effects?} To test whether our attack presents a concrete threat we performed a study on two crowdsourcing platforms.

We were interested in comparing the effectiveness of the phishing attacks. So we considered three hypotheses: For each two types of screenshots (authentic, green, plain) we tested whether one type is more trustworthy than the other, and did this for all combinations.

\begin{description}
\item[H1.] For a simple random sample, participants are likely to respond differently to a plain phishing attack site than to the original site. This hypothesis states the expectation that end-users can distinguish phishing sites from authentic sites if asked to. We use this hypothesis as motivation for the other hypotheses -- if a user cannot detect a phishing attack then improving upon the attack will not have a useful effect.
\item[H2.] For a simple random sample, participants are likely to respond differently to a green-bar attack site than to a plain attack site. This hypothesis states that end-users are more likely to fall for the green-bar attack.
\item[H3.] For a simple random sample, participants are likely to respond differently to a green-bar attack site than to the original site. This hypothesis states that end-users are still likely to detect green-bar attacks.
\end{description}
\newcommand{\hypothesis}[1]{\textbf{H#1}}

Note that the hypotheses are two-tailed and we do not test for direction.


While creating the sample screenshots we determined the following control variables:

\begin{itemize}
\item URL (between phishing attacks)
\item Page body
\item System clock 
\item Carrier
\item Wi-Fi signal strength
\item Battery level
\end{itemize}

To study the above hypotheses we made the following assumptions:

\begin{itemize}
\item We are able to select users at random from a general population.
\item The decision whether a given user is willing to enter personal information on a given web page depends only on what is displayed, i.e. is a function of the screen.
\item Users are able to accurately predict their decision for a given screen from only a screenshot.
\end{itemize}

We concluded from this that for a fixed web page, the decision for a randomly picked user whether or not to enter their personal information is distributed i.i.d. (independent and identically distributed) Bernoulli and is a function of the selected user.

\subsection{Particpants}

We recruited 3001 participants from Google's Customer Surveys platform, and 73 participants from CrowdFlower's platform. Each participant was paid depending on the platform. CrowdFlower workers were paid up to \$~0.20, and Google participants earned various benefits on the Google platform. 

For the CrowdFlower study participants were randomly split between 6 groups. Each group was shown one of the screenshots in Figure \ref{fig:survey-screenshots}, and participants were also presented with an anti-spam question. We also collected various background statistics about participants such as geo-location information.

We ran the intitial study on CrowdFlower with the intention of having 600 participants. However, we had a very small number of respondents per hour. Then we ran the study on Google Consumer Surveys with 3000 participants. After the results were in we started further exploratory surveys to investigate effects that may have impacted the study. 


\subsection{Study design}



We used two crowdsourcing platforms for this study: CrowdFlower and Google Consumer Surveys. As these platforms placed different restraints on the survey we had a different setup for each.



\subsubsection{CrowdFlower}

The CrowdFlower screen contained the following description:
\begin{quote}
Image Categorization

Let us know whether this website is safe based on the screenshot provided.

Thank You! Your careful attention on this task is greatly appreciated!
\end{quote}

We showed the screenshot and asked the following question:

\begin{quote}
Is this website safe?
\end{quote}

Available answers were ``Yes'' and ``No.'' In order to counter bias, we randomised the order of the answers through a JavaScript function in the CrowdFlower job.The CrowdFlower pa randomized the order of the answers to counter bias. 


As alternatives for wording the question we considered among others:

\begin{itemize}
\item Do you trust this site?
\item Would you login at this site?
\end{itemize}

However, each of these places inappropriate emphasis on aspects that are not related to connection security. For example, site trust would be significantly influenced by brand awareness. A user would also be more likely to respond positively to the login question if there is a prior relationship and the user has an account at that site with which to login.


To filter out automated (bot) submissions and ensure a minimum level of participant attentiveness, we asked a trivial control question:

\begin{quote}
What day of the week is it?
\end{quote}

We validated the response against a list of reasonable strings. We did not deem it necessary verify that the system-recorded timestamp corresponded to the entered weekday.

In addition to the questions that participants were asked to fill out, the CrowdFlower system automatically collected start/finish times for us. This allowed us to do a spot sanity check on time taken.

The CrowdFlower system automatically prevents multiple submissions from a single participant under the configuration that we had set up. We were not able to filter out participants from previous trial runs so we simply discarded their entries using the unique identifier. 

\begin{figure}
\centering
\frame{\includegraphics[width=\columnwidth]{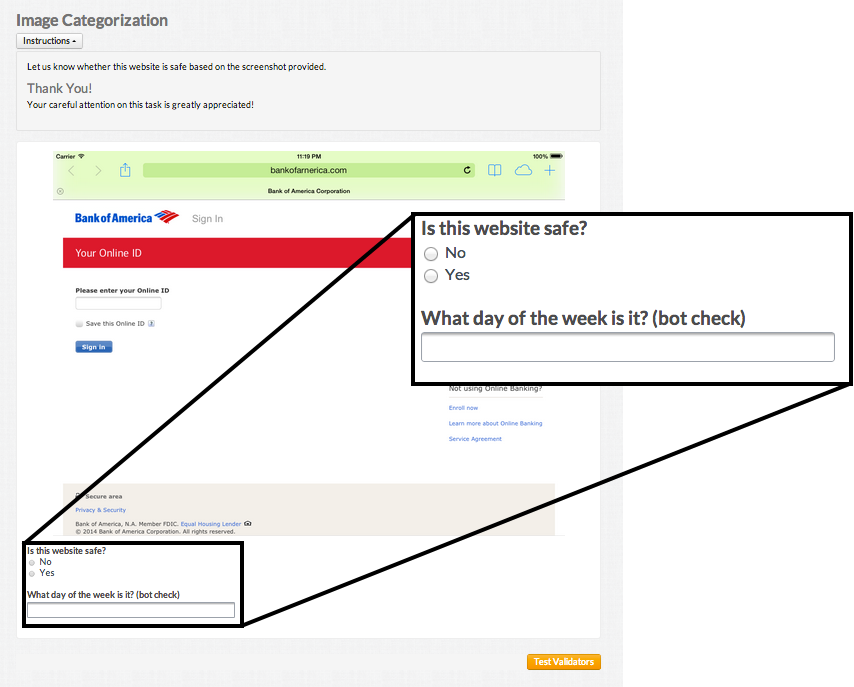}}
\caption{The CrowdFlower survey page with questions}
\label{fig:survey_sample}
\end{figure}

\begin{figure}
\centering
\frame{\includegraphics[width=0.45\columnwidth]{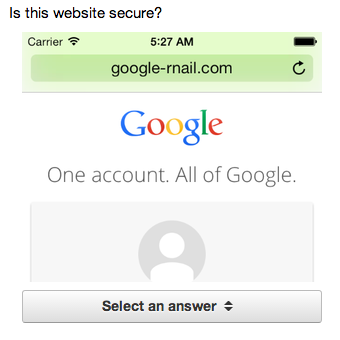}}
\frame{\includegraphics[width=0.45\columnwidth]{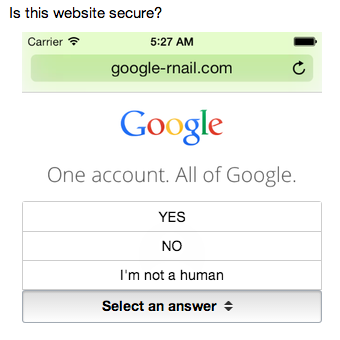}}
\caption{The Google Consumers Survey}
\label{fig:survey_google}
\end{figure}


\subsubsection{Google Consumer Surveys}

\begin{figure}
\centering
{
\begin{tabular}{rcc}
  & \textbf{Bank of America} & \textbf{Google Mail} \\\\
  & \emph{Authentic with EV-SSL} & \emph{Authentic with SSL}\\
  & \emph{bankofamerica.com} & \emph{accounts.google.com}\\
  & \frame{\includegraphics[width=0.2\textwidth]{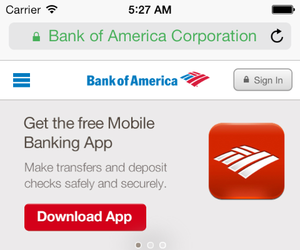}}
  & \frame{\includegraphics[width=0.2\textwidth]{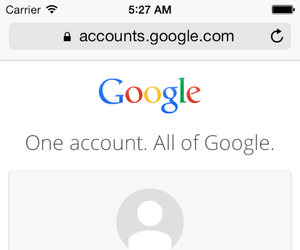}}\\\\
  & \emph{Fraud with green bar} & \emph{Fraud with green bar}\\
  & \emph{bankofarnerica.com} & \emph{google-rnail.com}\\
  & \frame{\includegraphics[width=0.2\textwidth]{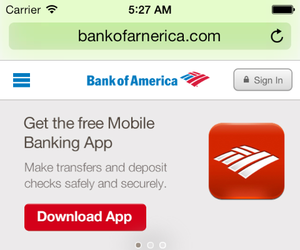}}
  & \frame{\includegraphics[width=0.2\textwidth]{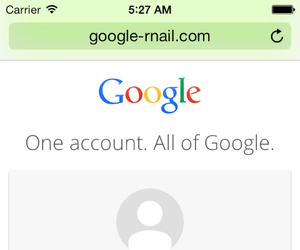}}\\\\
  & \emph{Fraud without green bar} & \emph{Fraud without green bar}\\
  & \emph{bankofarnerica.com} & \emph{google-rnail.com}\\
  & \frame{\includegraphics[width=0.2\textwidth]{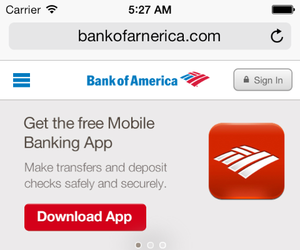}}
  & \frame{\includegraphics[width=0.2\textwidth]{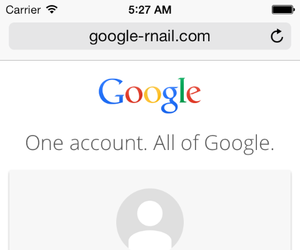}}
\end{tabular}
}
\caption{iPhone screenshots used in the survey}
\label{fig:survey-screenshots-iphone}
\end{figure}

For Google Consumer Surveys we adapted the CrowdFlower job, due to several technical and economic constraints. Google Consumer Surveys limited image size to 300 by 250 pixels. Hence we created iPhone screenshots instead of iPad screenshots and cut them off at the proportionate level. See figure \ref{fig:survey-screenshots-iphone} for these screenshots. Due to an oversight the question on the Google platform was ``Is this website secure?" instead of ``Is this website safe?".

Google Consumer Surveys also required significant extra payment per response to ask a second question (1000~\% over the price for one question). We thus did not include a bot control question and instead included an extra answer option: ``I'm not a human'' in the Yes/No question.

Finally, Google Consumer Surveys did not allow for a survey title nor description.

See the Section \ref{open_access} for more details about the materials used in this study.

\subsection{Results}

The data that we collected are summarised in Tables \ref{fig:results-google}, \ref{fig:results-google:proportions}, \ref{fig:results-cf:population}, and \ref{fig:results-cf:trusters}. We did not analyse the CrowdFlower data due to low volume. Note that we do not discuss the exploratory questions in this section, but that we will discuss these later.

\begin{table*}[ht]
\begin{center}
\caption{Google Consumer Survey responses, per category}
\mbox{}\\
\begin{tabular}{@{}l@{\quad}ll@{\quad}lllllll@{\quad}llllll@{\quad}llll@{}}
\toprule
                     & & & \multicolumn{4}{l}{B-of-A} & & & \multicolumn{4}{l}{Gmail} & & & \multicolumn{4}{l}{$\sum$} \\
\midrule
                     & & & Y    & N    & \eps& Pop & & & Y    & N    & \eps & Pop  & & & Y    & N    & \eps & Pop      \\
\midrule
Authentic            & & & 208  & 242  & 53  & 503 & & & 286  & 152  & 62   & 500  & & & 494  & 394  & 115  & 1003  \\
Fraud with green bar & & & 151  & 301  & 49  & 501 & & & 191  & 242  & 63   & 496  & & & 342  & 543  & 112  & 997   \\
Fraud with plain bar & & & 153  & 296  & 51  & 500 & & & 182  & 261  & 58   & 501  & & & 335  & 557  & 109  & 1001  \\
$\sum$               & & & 512  & 839  & 153 & 1504& & & 659  & 655  & 183  & 1497 & & & 1171 & 1494  & 336  & 3001  \\
\bottomrule
\end{tabular}
\label{fig:results-google}
\end{center}
\end{table*}




\begin{table}
\centering
\caption{Corrected sample proportions, per category}
\mbox{}\\
\begin{tabular}{|r|l|l|l|}
\hline
                     & B-of-A  & Gmail & Overall \\
\hline
Authentic            & 0.45    & 0.71  & 0.58    \\
\hline
Fraud with green bar & 0.29    & 0.42  & 0.35    \\
\hline
Fraud with plain bar & 0.29    & 0.38  & 0.34    \\
\hline
Overall              & 0.34    & 0.50  & 0.42    \\
\hline
\end{tabular}
\label{fig:results-google:proportions}
\end{table}

\begin{table}
\centering
\caption{Sample population, per category}
\mbox{}\\
\begin{tabular}{|r|l|l|l|}
\hline
\emph{CrowdFlower}   & B-of-A   & Google Mail   & $\sum$ \\
\hline
Authentic            & 9        & 10            & 19 \\
\hline
Fraud with green bar & 18       & 13            & 31 \\
\hline
Fraud with plain bar & 7        & 16            & 23 \\
\hline
$\sum$               & 34       & 39            & 73 \\
\hline
\end{tabular}
\label{fig:results-cf:population}
\end{table}

\begin{table}
\centering
\caption{Response of ``secure'', per category}
\mbox{}\\
\begin{tabular}{|r|l|l|l|}
\hline
\emph{CrowdFlower}   & B-of-A   & Google Mail   & $\sum$ \\
\hline
Authentic            & 7        & 9             & 16 \\
\hline
Fraud with green bar & 11       & 7             & 18 \\
\hline
Fraud with plain bar & 7        & 14            & 21 \\
\hline
$\sum$               & 25       & 30            & 55 \\
\hline
\end{tabular}
\label{fig:results-cf:trusters}
\end{table}


We tested the null hypotheses that there is pairwise no difference in trustworthiness. To do this we totalled the data over the sites Google Mail and Bank of America. We then ran Fisher's exact test and applied a Bonferri correction (factor 3) to account for family-wise error. The two-tailed p-values are summarized in Table \ref{fig:results:pvalues}.

By making additional assumptions on the Google Consumer Survey responses we were able to apply a correction to account for the bot responses. Specifically we assumed that bots (and unattentive human participants) chose at random between the possibilities. That is, bot responses are distributed i.i.d. $\mathcal U\{Y,N,\varepsilon\}$ where $\mathcal U$ is the uniform distribution.


Under these assumptions the best estimate for the human responses when we measured $(Y,N,\varepsilon)$ was:

\[ (\hat Y',\hat N')=(Y-\varepsilon,N-\varepsilon) \]

So we subtracted the ``I'm not a human'' response count from the ``Yes'' count and its triple from the sample population prior to summing across the sites and running the statistical tests.

The test we used is not exact because of this procedure, but it works in an approximative fashion,
for a signal-to-noise ratio approaching one (for $(Y+N)/(Y+N+\varepsilon)$ small).

\begin{table}
\centering
\caption{Corrected two-tailed p-values, by pair}
\mbox{}\\
\begin{tabular}{|r|l|l|l|}
\hline
                    & Green     & Authentic             & Authentic \\
                    & versus    & versus                & versus    \\
                    & Plain     & Green                 & Plain     \\
\hline
p-value             & 0.64    & $<10^{-15}$ & $<10^{-15}$ \\
\hline
Corrected           & $\ge 1$   & $<10^{-14}$ & $<10^{-14}$ \\
\hline
\end{tabular}
\label{fig:results:pvalues}
\end{table}

For a descriptive view of the data, we computed the confidence intervals at $\alpha = 0.05$ and $\alpha = 0.01$ for the sample proportion per type. We used the normal approximation which was appropriate because $Y, N \gg 5$ in all cases. These intervals as well as the observed sample proportions are shown in Figure \ref{fig:results:cintervals}.

\begin{table}
\centering
\caption{Confidence intervals, by type}
\mbox{}\\
\begin{tabular}{|r|l|l|l|l|l|}
\hline
            & $u(X)$    & $u(X)$    & $\hat{p}$ & $v(X)$    & $v(X)$ \\
$\alpha$    & 0.01      & 0.05      &           & 0.05      & 0.01 \\
\hline
Authentic   & 0.51      & 0.54      & 0.58      & 0.61      & 0.63 \\
\hline
Green       & 0.30      & 0.31      & 0.35      & 0.38      & 0.40 \\
\hline
Plain       & 0.29      & 0.30      & 0.34      & 0.37      & 0.38 \\
\hline
\end{tabular}
\label{fig:results:cintervals}
\end{table}

We also calculated effect size in the form of a pairwise odds ratio. This is shown in Figure \ref{fig:results:oddsratios}.

\begin{table}
\centering
\caption{Odds ratio, by pair}
\mbox{}\\
\begin{tabular}{|r|l|l|}
\hline
            & Green             & Plain \\
\hline
Authentic   & 2.54              & 2.69 \\
\hline
Green       & \cellcolor{gray}  & 1.06 \\
\hline
\end{tabular}
\label{fig:results:oddsratios}
\end{table}







\section{Discussion of Threat Level}


\subsection{Hypothesis Tests}

The survey data support \hypothesis 1 with statistical significance. We can then conclude that end-users can in fact detect phishing when they are asked about site security. The study of more effective phishing attacks is therefore motivated.
We were also able to confirm \hypothesis 3 which states that end-users can still detect the underlay green-bar phishing attack. 
While we did not test direction in the statistical tests the sample proportions and confidence intervals clearly show the direction of the effects.

Unfortunately we were not able to confirm \hypothesis 2 so it is not clear that our attack in this variant was effective. A larger sample size and/or an improved attack with larger effect size would be necessary to confirm this. The sample odds ratio was only 1.06 (with a 95~\% confidence interval of 0.84 to 1.34) so the attack does not appear to be a significant improvement in this variant.

Our results warrant further investigation. We have started several exploratory studies to quantitatively measure the effect of the many aspects in which the attack can be varied. Section \ref{sect:exploratory} goes into more detail of preliminary findings of these investigations.

\subsection{Interpretation of Results}

We suspect that in the context of the survey the green colouring may have drawn the participant's attention to the URL. Participants may have interpreted the green colouring as a highlight or emphasis, which may have pointed them to the URL bar as an area to pay attention to and base their decision on. We may thus have caused participants to read the URL, when they normally would not have done so even in the plain phishing screenshot. This would seem unlikely to happen during typical web browsing.


Previous studies by other authors have shown that end-users tend not to look at the address bar to verify that they are browsing a trusted website \cite{emperor}. Then the purpose of testing an alternative presentation might be unclear. However instead of simply subverting the original indicator we are in fact attempting to create a false indicator that could hint that the website in question is safe. While the end-user might not consciously recognize the indicator there is a chance that the user subconsciously responds to it.

One way to explain our results is that the human mind functions as a prediction engine. It might be that users have certain expectations, and if something is suddenly there that wasn't there before,  or if it doesn't serve as an affirmative signal, it will serve as an attention trigger. This indicates that salience of trust indicators may play a role, i.e. only those that are actually seen by the user will be considered. For phishers this indicates that they shouldn't draw attention if they don't have to, and for engineers of browsers that they should do more to make security (and insecurity) indicators more salient. These and various improvements to our attack are possible, and we discuss possible future work in Section \ref{chap:conclusion}. 

Note that in our survey we overlooked some variables before running the study such as the lock icon to indicate a secured connection. This may have adversely affected our results. In the future a different methodology might be appropriate such as comparing the screenshots using an ``image diff'' software to verify that no variables are overlooked.

\subsection{Exploratory Investigations} \label{sect:exploratory}

\begin{figure}
\centering
\frame{\includegraphics[width=0.45\columnwidth]{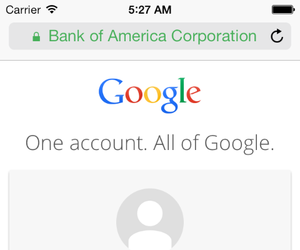}}
\caption{Mismatching EV screenshot}
\label{fig:variant:wrong_ev}
\end{figure}

\begin{figure}
\centering
\frame{\includegraphics[width=0.45\columnwidth]{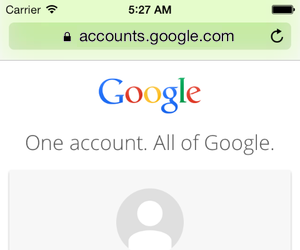}}
\frame{\includegraphics[width=0.45\columnwidth]{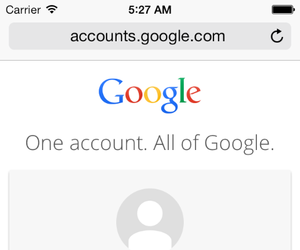}}
\caption{Variants of the authentic screenshot}
\label{fig:variant:authentic}
\end{figure}

After the primary study we ran several short runs of adapted interfaces. This exploratory investigation looked at the effect of the question we asked as well as several factors on the authentic screenshots. We performed the following short surveys:

\begin{description}

\item[``Safe'']
We tested the green and plain attacks but asked ``Is this website safe?'' This was to explore the effect of a slightly different question.

\item[Mismatching EV]
We presented a screenshot with the Bank of America EV but the Google Mail content to test whether users actually read the EV information or only notice whether an EV is displayed. The screenshot we used is shown in Figure \ref{fig:variant:wrong_ev}.

\item[Authentic without lock]
We showed the authentic screenshot but without lock. This tests the effect of the padlock icon. The screenshot can be seen in Figure \ref{fig:variant:authentic}.

\item[Authentic with green bar]
Similar to the underlay green-bar attack we presented the authentic site with a green address bar. This was to complement our original survey results. This screenshot is also included in Figure \ref{fig:variant:authentic}.

\end{description}

The survey results are shown in Table \ref{fig:results:explorative}.

Varying the question did not appear to have any effect. We observed sample proportions of 0.40 vs. 0.42 (odds ratio: 1.09) and 0.36 vs. 0.38 (OR: 1.09).

The mismatching EV on Google Mail appears to be noticeable enough (0.52 vs. 0.71, OR: 2.26). However it performs better than the green attack (0.42, OR: 1.50) and better than the plain attack (0.38, OR: 1.77). Interestingly the mismatching EV appears to perform better than the authentic Bank of America screenshot at 0.45 (OR: 1.32).

Of note, the green variant of the authentic Gmail website appears to perform worse than the original authentic screenshot (0.61 vs. 0.71, OR: 1.57). We suspect that this may be due to users recalling the look of the original authentic site on their own devices and noticing that it does not look the the same. The typical security indicators are present but this might not be enough to sway the user after they have noticed the difference.

Removing the lock from the authentic screenshot appears to have a great effect on user response. Without a padlock icon the screenshot performs at 0.48 compared to the original 0.71 (OR: 2.65).

\begin{table}
\centering
\caption{Results of the explorative surveys}
\mbox{}\\
\begin{tabular}{|r|l|l|l|l|l|l|}
\hline
                        & Y     & N     & \eps  & $\hat Y'$ & $\hat N'$ & $\hat p$ \\
\hline
Mismatching EV          & 83    & 79    & 25    & 58        & 54        & 0.52 \\
\hline
``Safe'' green          & 65    & 97    & 17    & 48        & 72        & 0.40 \\
\hline
``Safe'' plain          & 72    & 108   & 21    & 51        & 87        & 0.36 \\
\hline
Authentic green         & 86    & 66    & 32    & 54        & 34        & 0.61 \\
\hline
Authentic w/out lock    & 81    & 85    & 28    & 53        & 57        & 0.48 \\
\hline
\end{tabular}
\label{fig:results:explorative}
\end{table}

\subsection{Limitations}

Throughout the survey design and execution we encountered limitations both inherent in the survey method and induced by the specific crowd-sourcing platform that we used. Below we provide further details about these limitations.

Other researchers are encouraged to verify, as there are likely various hidden biases present in the study. We lower the barrier to others to verify our findings in different settings and using different methods as all our materials are made available under open access. We agree that there are still no widely agreed-upon standards for sharing study designs, so the study will need to be ported to the specific platform on which it is run.












\subsubsection{Representativeness of Materials}

An important aspect is the question of how representative the materials are. This is often a tougher question than sample representativeness, especially when only investigating the difference between treatments. For example: We haven't looked a safari's private browsing mode, which uses something like darkened glass effect. Additionally, we've looked at something like frosted glass, and it could be that there are interfaces that are much more transparent.

\subsubsection{Concerns around Quality of Responses}

We tried to keep the survey short and quick to complete for participants. Having only a single real question and a trivial bot control question does not necessarily ensure a high level of attentiveness. However, note that for investigations into phishing the application of an attentiveness check to filter participants is likely to filter out those that are more likely to fall for phishing.

Google Consumer Surveys showed a non-negligible level of responses with ``I'm not a human.'' We were not able to determine whether this was due to bot submissions or lack of attentiveness. The problem of distinguishing bots from non-attentive humans is a general problem in crowdsourced research. For future work we intend to run lab studies alongside crowdsourced studies. Due to an increased noise floor, there is a need to scale up the number of respondents in crowdsourced studies. In addition there is also the requirement to scale sideways: to run studies on multiple platforms, and preferably to run studies both on- and offline.

With only a Boolean yes/no response it is hard to tell whether people have correctly interpreted the question. The satisfaction survey suggests that the instructions were considered clear by participants. However, it is doubtful that a misunderstanding would be detected by enough participants. Note that the answers in the bot question on CrowdFlower indicate that the participants there read the questions, but that this does not say anything about the clarity and interpretation of the question about website trustworthiness. In future work a sanity check in the form of a dedicated attention-check question is appropriate on the Google Consumer Surveys platform. Additionally, open-ended answers asking how a question is interpreted are appropriate.






Unfortunately it is a common technique on some crowd-sourcing sites to perform automated submissions for personal gain. This leaves survey authors with invalid or useless submissions. To counter this we implemented a bot check. Unfortunately this bot check can filter out real participants if they are not attentive enough or simply by chance due to mismatching answer formats.

Google Consumer Surveys did not allow us to ask a bot control question as described earlier.

\subsubsection{Concerns around Technology}

CrowdFlower doesn't directly support surveys split into groups or with randomised answer ordering. We had to implement this using client-side JavaScript code. Then for choosing a random screenshot to show to the participant we had to rely on the browser's built-in random number generator. This may have impacted on random assignment into groups. 
Additionally, CrowdFlower didn't allow us to prevent duplicates across test runs vs. actual runs and this means we had to take out around a fifth of the final data.
Lastly, as shown by Renkema et al \cite{renkema2014buildling}, we cannot be sure that CrowdFlower is running their infrastructure properly (or any other crowdsourcing provider for that matter).

Google's platform did not allow us to start several surveys simultaneously and did not allow random assignment within a survey. 
In part this was due to an intransparent survey review process that introduced delay between the time of survey scheduling and the actual starting time. As time of day among other things can affect the participants' attentiveness this may have unnecessarily biased some groups and affected the outcome. Another possible source of bias is adaptive targeting (whereby Google Consumer Surveys appears to have directed the surveys towards different audiences), which likely introduced further unknown bias. 

Generally, the crowdsourcing platforms are very opague. The review process for Google Consumer Surveys was also intransparent with respect to which surveys they allowed and which they didn't. In one instance the exact same survey with a slightly adapted image was rejected even though ten other similar surveys were accepted. After making a change with no effect and resubmitting the survey it was accepted.









\subsection{Ethics}

During the data collection various demographics were automatically included by Google. For CrowdFlower we enabled the option of saving the browser agent identifier as well as the IP address. We chose to save these to enable checking of double submissions. For opening of the data (see Section \ref{open_access}) we have opted to remove the IP addresses and to anonymise the user IDs for both platforms. We expect the participant terms of use of these crowdsourcing platforms to account for this.

There was no informed consent from the participants. However, we assumed implicit consent for non-harmful tasks, given that the participants working on a crowdsourcing platform. Our surveys did not provide a debrief informing the participants of the purpose of the study, and as such participants were not made aware of deception. This was largely due to technical limitations in the platforms available to us, as well as the nature of crowdsourcing. Given the nature of crowdsourcing, participants are not likely to read any debrief text.

We think that there is no likely harm from participating in the study, and it might even help people in becoming slightly more aware of web security. Also, participants in the CrowdFlower platform were able to take part in a satisfaction survey to rate their experience with the original survey. We do not have access to the raw survey data but the summary statistics show a satisfaction level of at least 4 out of 5 in the categories ``Overall'', ``Instructions Clear'', ``Ease Of Job'' and ``Pay.''






\section{Proposed Solution}\label{chap:solution}


In the interest of responsible disclosure we make Apple aware of the problem after we carried out the user study and before publication, and recommend them to remove transparency in iOS7 Safari. While this will take away some of the ``eye candy" of the interface, it will ensure that the attack presented in this paper is no longer possible. Alternatively, on first displaying a new page, the URL bar could be made opaque and only becoming transparent after some time. This would not provide complete protection, but it could provide an acceptable compromise between seduction and security.

Additionally we encourage all browser vendors, for both desktop and mobile web browsers, to come together and agree on common standards for displaying trust indicators. Standards for secure interfaces are sorely lacking. We expect that one factor that contributes to the difficulty of distinguishing a fake from a genuine trust indicator is the current wide variety of trust signals. 

\section{Related Work}

Security systems are known to fail due to human factors \cite{fail}. This is the same for phishing \cite{Dhamija:2006:WPW:1124772.1124861}, and this is what our attack builds on. We have presented a hybrid attack that combines both technical shortcomings with limitations in human capabilities.

As mentioned by Yee \cite{yee2005guidelines}, user interface elements should not provide control to an attacker. The attack that we have discovered does just this: we achieve limited control of the user interface to influence the way that trust indicators are displayed. What is enabled by our approach is twofold: spoofing security indicators, and making insecurity indicators harder to perceive. Following an econometric analysis, Herley \cite{externalities} suggests that users don't check security indicators because the real costs are greater than possibly damages. While users would be unlikely to make such extensive calculations, we can safely assume that they make some cost-benefit analysis when exposed to various technologies. We use this propensity in our attack scenario, by seeking to make checking of security indicators as difficult as possible.

Conti et al \cite{overload} describe attacks against security systems based on visual overload, and also describe the need for countermeasures against malicious interfaces \cite{conti2010malicious}. The literature describes various malicous interface methods that go in a similar direction to ours. Related attacks are clickjacking \cite{huang2012clickjacking}, UI redressing, and mousejacking.

An attack that also depends on relocation of the page is the URL spoofing attack from Dhanjani \cite{iphone}. Instead of trying to change the authentic URL, Dhanjani creates a visually identical copy of the URL bar in the iPhone interface. Gabrilovich and Gontmakher \cite{gabrilovich2002homograph} describe a way of spoofing characters using look-alike glyphs from the Unicode character set. Krammer \cite{Krammer:2006:PDA:1501434.1501473} describes a possible countermeasure for this based on highlighting. In transluscent interfaces such highlighting might be circumvented using the underlay attack that we have presented.

Lin et al \cite{lin2007} have described URL highlighting as a mechanism that has a protective effect in certain cases. Again, our technique might be able to reduce it's effectiveness when transparent interfaces are present. Another countermeasures that depends on visualisation are visual fingerprints \cite{dhamija2005battle} based on hash visualisation \cite{visualisation}, which is also possible circumvented by our underlay attack.

Attacks such as those presented here illustrate the need for trusted windowing systems such as EROS by Shapiro et al \cite{shapiro2004design}. However, this is not likely to happen in the near future. Others have shown the sorry state of security indicators and security warnings on the web. Stebila \cite{misuse} found many websites abusing trust indicators. Schechter et al \cite{emperor} found that security indicators were generally ineffective. Jakobsson \cite{indicators} found that padlock indicators had little effect on trust. Amrutkar et al \cite{amrutkar2012measuring} performed a usability analysis of indicators in web browsers, and found them severly lacking. 




\section{Conclusion}\label{chap:conclusion}

We have highlighted the danger of transparency in user interfaces, and have presented an attack on iOS7 Safari. We have conducted an evaluation of the attack on a crowdsourcing platform, and we found that further investigation is needed.

Various avenues for future work are open:
\begin{itemize}
\item Testing which colour works best. An attacker could do this adaptively using, for example, A/B testing \cite{kohavi2009controlled} or Bandit based methods \cite{white2012bandit}.
\item Testing phishing URLs by running them through a crowdsourcing survey platform.
\item Testing phishing email content using crowdsourcing.
\item Investigating whether the ideal colour is site-specific.
\item Investigating the possibility of defeating the blurring mask to present security indicators.
\item Making phishing URL harder to distinguish, e.g by camouflaging the misspellings using reduced contrast, distracting patterns or by derailing attention towards less suspicious UI elements
\end{itemize}

Through these avenues we can explore opportunities for user confusion. Bravo-Lillo et al \cite{{Bravo-Lillo:2013:YAP:2501604.2501610}} have advocated the design of ``attractors" in order to direct the user's attention towards relevant aspects of security warnings. We propose further research into the design of ``distractors" that can hinder users in interpreting security indicators.





\section{Open Access}\label{open_access}


\emph{Note to reviewers: The data will be added in the conference version due to anonymity issues around file authorship.}




\bibliographystyle{IEEEtran}
\bibliography{references}

\begin{thebibliography}{10}
\providecommand{\url}[1]{#1}
\csname url@samestyle\endcsname
\providecommand{\newblock}{\relax}
\providecommand{\bibinfo}[2]{#2}
\providecommand{\BIBentrySTDinterwordspacing}{\spaceskip=0pt\relax}
\providecommand{\BIBentryALTinterwordstretchfactor}{4}
\providecommand{\BIBentryALTinterwordspacing}{\spaceskip=\fontdimen2\font plus
\BIBentryALTinterwordstretchfactor\fontdimen3\font minus
  \fontdimen4\font\relax}
\providecommand{\BIBforeignlanguage}[2]{{%
\expandafter\ifx\csname l@#1\endcsname\relax
\typeout{** WARNING: IEEEtran.bst: No hyphenation pattern has been}%
\typeout{** loaded for the language `#1'. Using the pattern for}%
\typeout{** the default language instead.}%
\else
\language=\csname l@#1\endcsname
\fi
#2}}
\providecommand{\BIBdecl}{\relax}
\BIBdecl

\bibitem{yee2005guidelines}
K.-P. Yee, ``Guidelines and strategies for secure interaction design,''
  \emph{Security and Usability: Designing Secure Systems That People Can Use},
  pp. 247--273, 2005.

\bibitem{emperor}
S.~E. Schechter, R.~Dhamija, A.~Ozment, and I.~Fischer, ``{The Emperor's New
  Security Indicators},'' in \emph{2007 IEEE Symposium on Security and
  Privacy}, Oakland, 2007, pp. 51--65.

\bibitem{renkema2014buildling}
A.~Renkema-Padmos, M.~Volkamer, and K.~Renaud, ``Building castles in quicksand:
  Blueprint for a crowdsourced study,'' in \emph{CHI'14 Extended Abstracts on
  Human Factors in Computing Systems}.\hskip 1em plus 0.5em minus 0.4em\relax
  ACM, 2014.

\bibitem{fail}
R.~Anderson, ``{Why Cryptosystems Fail},'' in \emph{1st ACM Conference on
  Computer and Communications Security}, Fairfax, 1993, pp. 215--227.

\bibitem{Dhamija:2006:WPW:1124772.1124861}
\BIBentryALTinterwordspacing
R.~Dhamija, J.~D. Tygar, and M.~Hearst, ``Why phishing works,'' in
  \emph{Proceedings of the SIGCHI Conference on Human Factors in Computing
  Systems}, ser. CHI '06.\hskip 1em plus 0.5em minus 0.4em\relax New York, NY,
  USA: ACM, 2006, pp. 581--590. [Online]. Available:
  \url{http://doi.acm.org/10.1145/1124772.1124861}
\BIBentrySTDinterwordspacing

\bibitem{externalities}
C.~Herley, ``{So Long, and No Thanks for the Externalities: The Rational
  Rejection of Security Advice by Users},'' in \emph{2009 Workshop on New
  Security Paradigms}, Oxford, 2009, pp. 133--144.

\bibitem{overload}
G.~Conti, M.~Ahamad, and J.~Stasko, ``{Attacking Information Visualization
  System Usability Overloading and Deceiving the Human},'' in \emph{Symposium
  On Usable Privacy and Security 2005}, Pittsburgh, 2005, pp. 89--100.

\bibitem{conti2010malicious}
G.~Conti and E.~Sobiesk, ``Malicious interface design: exploiting the user,''
  in \emph{Proceedings of the 19th international conference on World wide
  web}.\hskip 1em plus 0.5em minus 0.4em\relax ACM, 2010, pp. 271--280.

\bibitem{huang2012clickjacking}
L.-S. Huang, A.~Moshchuk, H.~J. Wang, S.~Schechter, and C.~Jackson,
  ``Clickjacking: attacks and defenses,'' in \emph{Proceedings of the 21st
  USENIX Security Symposium}, 2012.

\bibitem{iphone}
N.~Dhanjani, ``{New Age Application Attacks Against Apple's iOS and
  Countermeasures},'' in \emph{Black Hat Europe 2011}, Barcelona, 2011.

\bibitem{gabrilovich2002homograph}
E.~Gabrilovich and A.~Gontmakher, ``The homograph attack,''
  \emph{Communications of the ACM}, vol.~45, no.~2, p. 128, 2002.

\bibitem{Krammer:2006:PDA:1501434.1501473}
\BIBentryALTinterwordspacing
V.~Krammer, ``Phishing defense against idn address spoofing attacks,'' in
  \emph{Proceedings of the 2006 International Conference on Privacy, Security
  and Trust: Bridge the Gap Between PST Technologies and Business Services},
  ser. PST '06.\hskip 1em plus 0.5em minus 0.4em\relax New York, NY, USA: ACM,
  2006, pp. 32:1--32:9. [Online]. Available:
  \url{http://doi.acm.org/10.1145/1501434.1501473}
\BIBentrySTDinterwordspacing

\bibitem{lin2007}
E.~Lin, S.~Greenberg, E.~Trotter, D.~Ma, and J.~Aycock, ``{Does Domain
  Highlighting Help People Identify Phishing Sites?}'' in \emph{CHI 2011
  Conference on Human Factors in Computing Systems}, Vancouver, 2011, pp.
  2075--2084.

\bibitem{dhamija2005battle}
R.~Dhamija and J.~D. Tygar, ``The battle against phishing: Dynamic security
  skins,'' in \emph{Proceedings of the 2005 symposium on Usable privacy and
  security}.\hskip 1em plus 0.5em minus 0.4em\relax ACM, 2005, pp. 77--88.

\bibitem{visualisation}
A.~Perrig and D.~Song, ``{Hash Visualisation: A New Technique to Improve
  Real-World Security},'' in \emph{International Workshop on Cryptographic
  Techniques and E-Commerce}, Hong Kong, 1999.

\bibitem{shapiro2004design}
J.~S. Shapiro, J.~Vanderburgh, E.~Northup, and D.~Chizmadia, ``Design of the
  eros trusted window system,'' in \emph{Proceedings of the 13th conference on
  USENIX Security Symposium-Volume 13}.\hskip 1em plus 0.5em minus 0.4em\relax
  USENIX Association, 2004, pp. 12--12.

\bibitem{misuse}
D.~Stebila, ``{Reinforcing Bad Behaviour: The Misuse of Security Indicators on
  Popular Websites},'' in \emph{OzCHI 2010}, Brisbane, 2010, pp. 248--251.

\bibitem{indicators}
M.~Jakobsson, A.~Tsow, A.~Shah, E.~Blevis, and Y.-K. Lim, ``{What Instills
  Trust? A Qualitative Study of Phishing},'' in \emph{11th International
  Conference on Financial cryptography and 1st International Conference on
  Usable Security}, Scarborough, 2007, pp. 356--361.

\bibitem{amrutkar2012measuring}
C.~Amrutkar, P.~Traynor, and P.~C. van Oorschot, ``Measuring ssl indicators on
  mobile browsers: extended life, or end of the road?'' in \emph{Information
  Security}.\hskip 1em plus 0.5em minus 0.4em\relax Springer, 2012, pp.
  86--103.

\bibitem{kohavi2009controlled}
\BIBentryALTinterwordspacing
R.~Kohavi, R.~Longbotham, D.~Sommerfield, and R.~Henne,
  ``\BIBforeignlanguage{English}{Controlled experiments on the web: survey and
  practical guide},'' \emph{\BIBforeignlanguage{English}{Data Mining and
  Knowledge Discovery}}, vol.~18, no.~1, pp. 140--181, 2009. [Online].
  Available: \url{http://dx.doi.org/10.1007/s10618-008-0114-1}
\BIBentrySTDinterwordspacing

\bibitem{white2012bandit}
J.~White, \emph{Bandit Algorithms for Website Optimization}.\hskip 1em plus
  0.5em minus 0.4em\relax " O'Reilly Media, Inc.", 2012.

\bibitem{Bravo-Lillo:2013:YAP:2501604.2501610}
\BIBentryALTinterwordspacing
C.~Bravo-Lillo, S.~Komanduri, L.~F. Cranor, R.~W. Reeder, M.~Sleeper, J.~Downs,
  and S.~Schechter, ``Your attention please: Designing security-decision uis to
  make genuine risks harder to ignore,'' in \emph{Proceedings of the Ninth
  Symposium on Usable Privacy and Security}, ser. SOUPS '13.\hskip 1em plus
  0.5em minus 0.4em\relax New York, NY, USA: ACM, 2013, pp. 6:1--6:12.
  [Online]. Available: \url{http://doi.acm.org/10.1145/2501604.2501610}
\BIBentrySTDinterwordspacing

\end{thebibliography}
\end{document}